\documentclass[fleqn,usenatbib]{mnras}

\usepackage[T1]{fontenc}

\DeclareRobustCommand{\VAN}[3]{#2}
\let\VANthebibliography\thebibliography
\def\thebibliography{\DeclareRobustCommand{\VAN}[3]{##3}\VANthebibliography}

\usepackage{graphicx}	% Including figure files
\usepackage{amsmath}	% Advanced maths commands
\usepackage{amssymb}	% Extra maths symbols
\usepackage{soul}
\usepackage[]{xcolor}

\usepackage{newtxtext,newtxmath}
\newcommand{\angstrom}{\mbox{\normalfont\AA}}
\newcommand{\hii}{\ifmmode \text{H}\,\textsc{ii} \else H~{\scshape ii}\fi}
\title[SMC Stellar Clusters with S-PLUS]{Ages and metallicities of stellar clusters using S-PLUS narrow-band integrated photometry: the Small Magellanic Cloud}

\author[G. Fabiano de Souza et al.]{G. Fabiano de Souza,$^{1}$\thanks{E-mail: gabriel.fabiano.souza@usp.br}
P. Westera,$^{2}$ %ufabc
F. Almeida-Fernandes,$^{1, 3}$
G. Limberg,$^{1, 4, 5}$
\newauthor
B. Dias,$^6$
J. A. Hernandez-Jimenez,$^{7}$
F. R. Herpich,$^{1}$
L. O. Kerber,$^{8}$ 
E. Machado-Pereira,$^{9}$
\newauthor
H. D. Perottoni,$^{1,10}$
Rafael Guerço,$^{9}$
L. Li,$^{1}$
L. Sampedro,$^{1}$
A. Kanaan,$^{11}$
T. Ribeiro,$^{12}$
\newauthor
W. Schoenell,$^{13}$
C. Mendes de Oliveira$^{1}$
\\
$^{1}$Universidade de S\~ao Paulo, Instituto de Astronomia, Geof\'isica e Ci\^encias Atmosf\'ericas, Departamento de Astronomia, SP 05508-090, S\~ao Paulo, Brazil\\
$^{2}$Universidade Federal do ABC (UFABC), Rua Santa Ad\'elia, 166, 09210-170, Santo Andr\'e - SP, Brazil\\
$^{3}$ NSF’s NOIRLab, 950 N. Cherry Ave., Tucson, AZ 85719, USA \\
$^{4}$ Department of Astronomy \& Astrophysics, University of Chicago, 5640 S. Ellis Avenue, Chicago, IL 60637, USA \\
$^{5}$ Kavli Institute for Cosmological Physics, University of Chicago, Chicago, IL 60637, USA \\
$^{6}$ Instituto de Astrofísica, Facultad de Ciencias Exactas, Universidad Andres Bello, Av. Fernandez Concha 700, 7591538, Las Condes, Santiago, Chile\\
$^{7}$ Universidade do Vale do Paraíba, Av. Shishima Hifumi, 2911, Zip Code 12244-000, São José dos Campos, SP, Brazil \\
$^{8}$Universidade Estadual de Santa Cruz, Depto. de Ci\^encias Exatas e Tecnol\'ogicas
Rodovia Jorge Amado km 16, 45662-900, Ilh\'eus, Brazil \\
$^{9}$ Observat\'orio Nacional, Rua General Jos\'e Cristino 77, CEP: 20921-400, S\~ao Crist\'ov\~ao, Rio de Janeiro, Brazil \\
$^{10}$ Nicolaus Copernicus Astronomical Center, Polish Academy of Sciences, ul. Bartycka 18, 00-716, Warsaw, Poland \\
$^{11}$ Departamento de F\'isica, Universidade Federal de Santa Catarina, Florianópolis, SC 88040-900, Brazil \\
$^{12}$ Rubin Observatory Project Office, 950 N. Cherry Ave., Tucson, AZ 85719, USA \\ 
$^{13}$ GMTO Corporation 465 N. Halstead Street, Suite 250 Pasadena, CA 91107, USA \\
}

\date{Accepted 2023 October 20. Received 2023 September 29; in original form 2023 May 18}

\pubyear{2024}

\begin{document}
\label{firstpage}
\pagerange{\pageref{firstpage}--\pageref{lastpage}}
\maketitle

\begin{abstract}

The Magellanic Clouds are the most massive and closest satellite galaxies of the Milky Way, with stars covering ages from a few Myr up to 13 Gyr. This makes them important for validating integrated light methods to study stellar populations and star-formation processes, which can be applied to more distant galaxies. We characterized a set of stellar clusters in the Small Magellanic Cloud (SMC), using the \textit{Southern Photometric Local Universe Survey}. This is the first age (metallicity) determination for 11 (65) clusters of this sample. Through its 7 narrow bands, centered on important spectral features, and 5 broad bands, we can retrieve detailed information about stellar populations. We obtained ages and metallicities for all stellar clusters using the Bayesian spectral energy distribution fitting code \texttt{BAGPIPES}. With a sample of clusters in the color range $-0.20 < r-z < +0.35$, for which our determined parameters are most reliable, we modeled the age-metallicity relation of SMC. At any given age, the metallicities of SMC clusters are lower than those of both the Gaia Sausage-Enceladus disrupted dwarf galaxy and the Milky Way. In comparison with literature values, differences are $\Delta$log(age)$\approx0.31$ and $\Delta$[Fe/H]$\approx0.41$, which is comparable to low-resolution spectroscopy of individual stars. Finally, we confirm a previously known gradient, with younger clusters in the center and older ones preferentially located in the outermost regions. On the other hand, we found no evidence of a significant metallicity gradient.

\end{abstract}

% Select between one and six entries from the list of approved keywords.
% Don't make up new ones.

%%%%%%%%%%%%%%%%%%%%%%%%%%%%%%%%%%%%%%%%%%%%%%%%%%

%%%%%%%%%%%%%%%%% BODY OF PAPER %%%%%%%%%%%%%%%%%%

\section {Introduction}
\label{introduction}

Populations of star clusters are excellent tools for understanding the formation and evolution of galaxies, as we can derive age, metallicity, and kinematics. Star clusters in distant galaxies can be studied using integrated light when it is not possible to resolve stars. Local Group galaxies can be studied in more detail, and be used as a laboratory to understand the regime of application of integrated light techniques that are used to analyze more distant galaxies. In this context, the Small Magellanic Cloud (SMC), is a satellite of the Milky Way (MW) that constitutes a valuable laboratory for calibrating stellar cluster characterization methods. Among the reasons that establish the importance of the SMC, we summarize a few:

\begin{itemize}

\item It is close enough \citep[$\sim$60\,kpc,][]{Gieren2018A&A...620A..99G} to have stellar content spatially resolved, allowing stellar parameter determination of individual stars;
\item It has hundreds of stellar clusters, covering ages ranging from a few million years to about 10 billion years, but with an age-metallicity relation different from the one found in the Milky Way \citep{Leaman2013MNRAS.436..122L, Parisi2022A&A...662A..75P};
\item In particular, the SMC has massive clusters (${>}10^4 M_{\odot}$), younger than 5 billion years and with sub-solar metallicities \citep[e.g. NGC 330,][]{UPB+17.2017MNRAS.468.3828U}, a combination non-existent in and, thus, complementary to the Galaxy;
\item The trio Large Magellanic Cloud (LMC), SMC, and MW is heavily interacting with each other and leaving traces in their stellar populations \citep[e.g.][]{Niederhofer2018A&A...613L...8N, Zivick2019ApJ...874...78Z, DeLeo2020MNRAS.495...98D, Dias2022MNRAS.512.4334D}.  
\item The SMC star cluster population has a growing number of objects fully characterized in terms of ages and metallicities with detailed analysis that serve as a reference for direct comparison and calibration of integrated-light based studies.

\end{itemize}

\par  For the last few decades, in several studies, ages from isochrone fitting in deep color-magnitude diagrams \citep[CMDs,][]{2007A&A...462..139K, 2014ApJ...797...35G,DKB+16.2016A&A...591A..11D, Milone2023A&A...672A.161M} and  metallicities from stellar spectroscopy  \citep[][]{PGC+14.2014AJ....147...71P,PGC+15.2015AJ....149..154P,PGC+16.2016AJ....152...58P, Song2021MNRAS.504.4160S, Parisi2022A&A...662A..75P, Dalessandro2016ApJ...829...77D, Mucciarelli2023A&A...677A..61M} have been determined. Besides, there are also integrated photometry studies of stellar clusters in the Magellanic Clouds \citep[MCs;][]{Hunter2003AJ....126.1836H, Goudfrooij2006MNRAS.369..697G, deGrijs2008MNRAS.383.1000D, Pandey2010MNRAS.403.1491P, Asa2012MNRAS.419.2116A, Popescu2012ApJ...751..122P}, allowing a compilation of a sample of MCs clusters with well-known properties. Such a sample is important for extragalactic studies since it allows accurate calibration of integrated light models of simple stellar populations based on spectra \citep{Gonzales2010MNRAS.403..797G, Asa2016MNRAS.457.2151A, Asa2022MNRAS.512.2014A, UPB+17.2017MNRAS.468.3828U, Goncalves2020MNRAS.499.2327G} and broadband photometry \citep{Popescu2012ApJ...751..122P}. In many cases, the age-metallicity degeneracy \citep{Worthey1994} is also present in integrated light. Although integrated spectroscopy and resolved photometry are capable of reducing the degeneracy, they are observationally much more expensive than techniques using integrated photometry. An intermediate path between the two approaches above can be provided by photometric studies based on a photometric system with narrow filters centered on important spectral characteristics that allow us to study these stellar populations. The path taken by this paper will be the use of the intermediate approach in order to determine ages and metallicities for a large sample of SMC clusters. 

The goal of this work is two-fold. The first is to use the SMC as a test case, for which we devised a technique to obtain cluster ages and metallicities from their integrated light, using multi-band photometry and assuming a proper calibration. The second goal is to provide first-time measurements of the ages and metallicities of a number of SMC clusters and analyze them as a population to understand the SMC chemical evolution. In the future, the same tool can be used for obtaining cluster parameters for more distant galaxies, where no resolved photometry will be available. To achieve our goals, we used integrated photometry from the {\it Southern Photometric Local Universe Survey} \citep[S-PLUS\footnote{\url{http://www.splus.iag.usp.br}}, discussed in Section \ref{subsec:splus},][]{MendesdeOliveira2019}.

We note that S-PLUS resolved CMDs for these SMC clusters, unfortunately, could not be used for double-checking the integrated photometry results. This is due to crowding in the cores of clusters, which makes resolved stars available only in their outskirts, limiting the number of sources usable for, e.g., an isochrone-fitting approach. Moreover, the photometric depth is not enough to reach the main sequence turnoff. However, we were able to validate our results with ages and metallicities from deep and resolved CMD from the Hubble Space Telescope \citep{Mighell1998AJ....116.2395M, Dolphin2001ApJ...562..303D} and the VISCACHA Survey \citep{Maia2019MNRAS.484.5702M} for some of our sample. Future analysis using spatially resolved deeper data from Rubin Observatory \citep{VeraRubin2020arXiv200907653V} may be able to check our results based on stellar cluster integrated photometry and may further validate the technique devised here for cases when spatially resolved photometry is not available.

The paper is organized as follows. In Section \ref{sec:photometry} we describe how the data is obtained. In Section \ref{sec:methods} we present the method used in this work. The results and discussions are in Section \ref{sec:results}, and conclusions are in Section \ref{conclusao}. 
\section{Integrated Photometry and Sample Selection}
\label{sec:photometry}

\subsection{S-PLUS}
\label{subsec:splus}

\begin{figure*}
\begin{center}
\includegraphics[width=2\columnwidth]{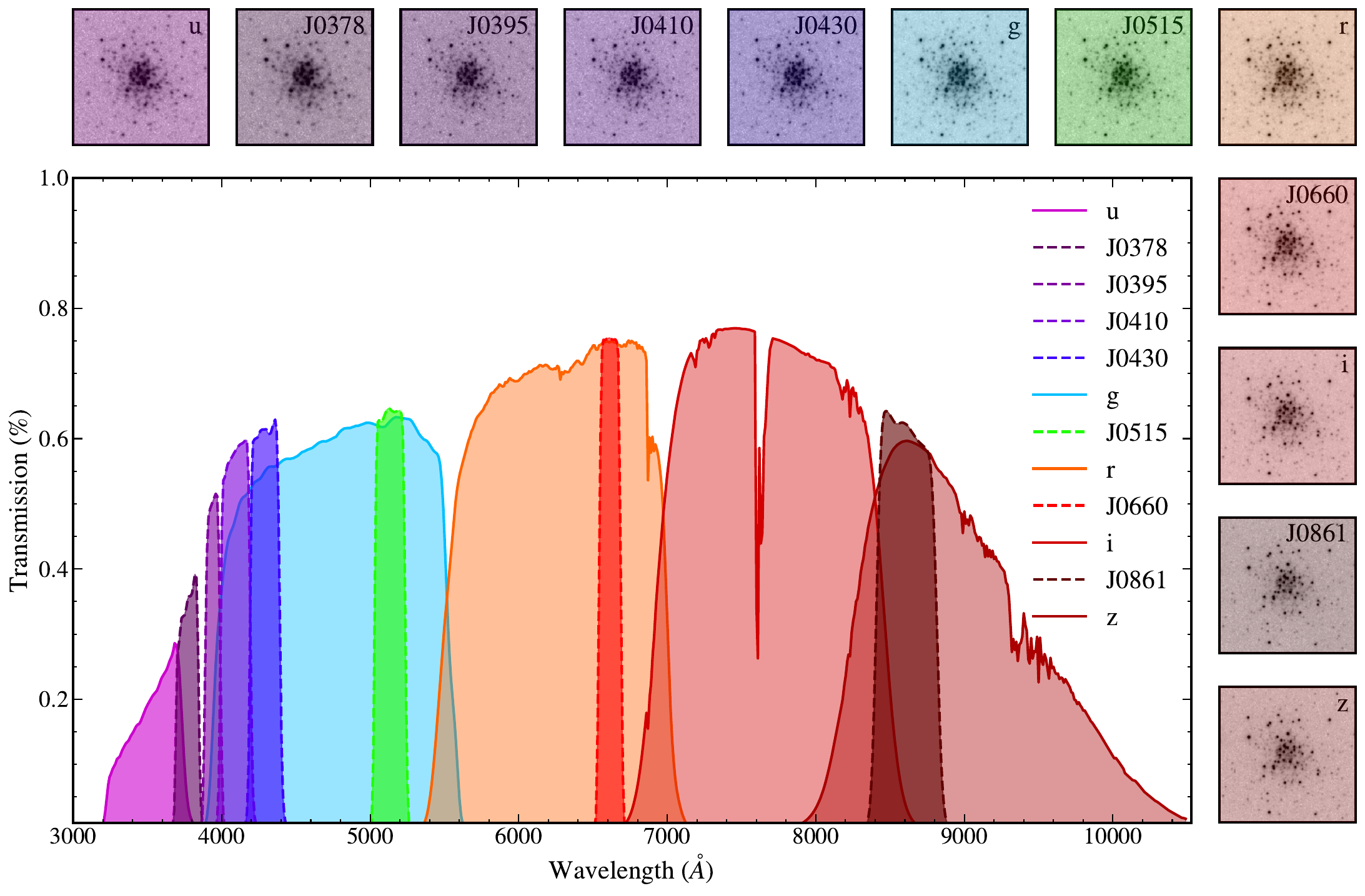}
\caption{Main panel: Transmission curves of the S-PLUS filters, based on laboratory measurements after considering the sky transmission curve, mirror reflectivity, and quantum efficiency of the detector \citep{MendesdeOliveira2019}. The panels above and to the right are the images of NGC 458 in the 12 filters.}
\label{fig:splus}
\end{center}
\end{figure*}

The S-PLUS is an astronomical mapping of the southern hemisphere which, when complete, will cover about $\sim$9300 square degrees of the sky. The S-PLUS uses the Javalambre system of 12 optical filters, five broad and seven narrow, designed by \citet{MarinFranch12} and originally used by the {\it Javalambre Photometric Local Universe Survey} \citep[J-PLUS\footnote{\url{http://www.j-plus.es/}},][]{Cenarro2019A&A...622A.176C}. This filter system represents a great opportunity to characterize the stellar clusters of the Magellanic Clouds. In Figure \ref{fig:splus} we present the transmission curves of the filters used in S-PLUS, together with the images of NGC 458 cluster in all 12 bands.

\begin{figure*}
\begin{center}
\includegraphics[width=2\columnwidth]{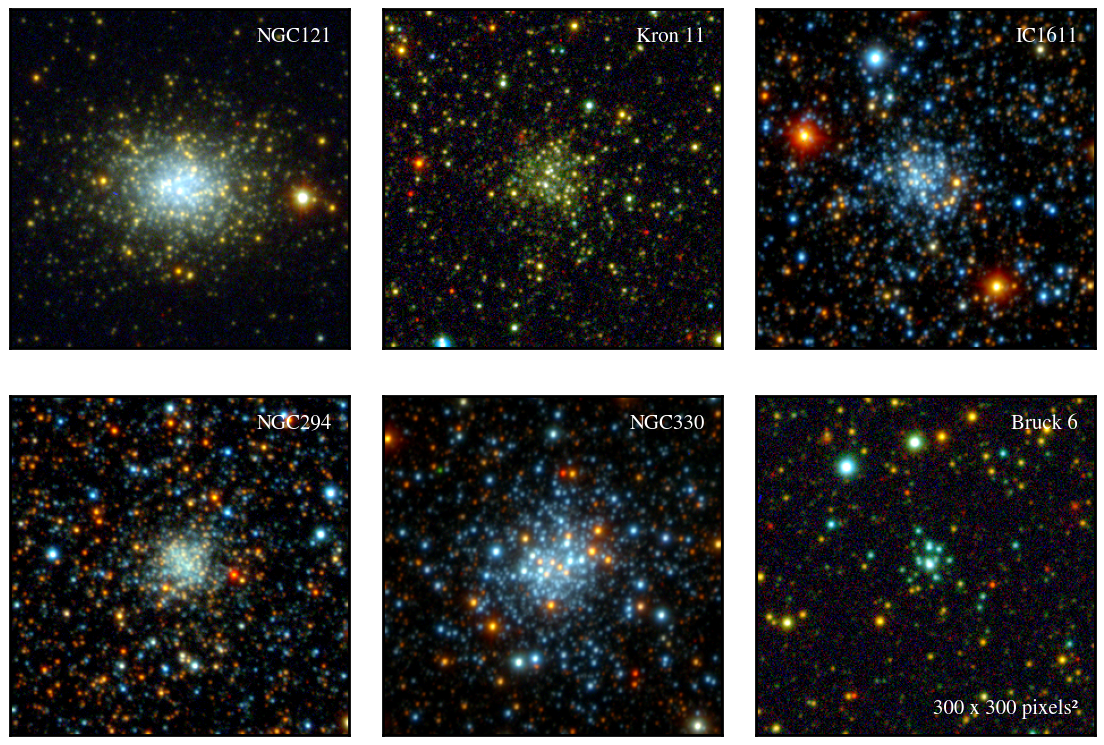}
\caption{Images of 6 stellar clusters studied in this work, made from the combination of the 12 bands using the \textit{S-PLUS cloud}, which can be found at \url{https://splus.cloud/}. The RGB was composed using the bands $r$, $i$, $J0861$, and $z$ for the R canal. For the G canal, we used the $g$, $J0515$, and $J0660$ bands. Finally, for the B canal, the  $u$, $J0378$, $J0395$, $J0410$, and $J0430$  were used.}
\label{fig:splus_smcclusters}
\end{center}
\end{figure*}

\begin{figure}
\begin{center}
\includegraphics[width=1\columnwidth]{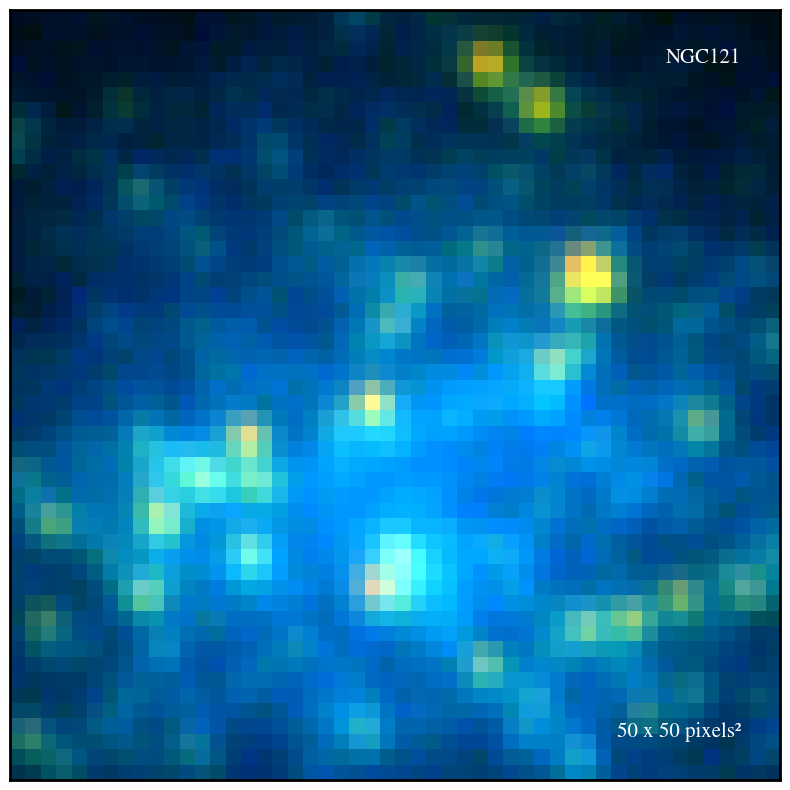}
\caption{Zoom of the stellar cluster NGC 121 showing that it is not possible to resolve stars in crowded core regions.}
\label{fig:splus_smcclusters_zoom}
\end{center}
\end{figure}

In order to study SMC star clusters, we analyzed images of $2$ deg$^2$ from S-PLUS, the only survey with multiple narrow-band filters covering this region of the sky. The SMC was observed with a typical \textit{seeing} of $1.5''$ - better than $2''$ in any case, air mass values between $1.4$ and $1.6$ and exposure times ranging from $3 \times 33$ to $3 \times 290$ seconds (depending on the filter), as described in Table 5 of \citet{MendesdeOliveira2019}. In Figure \ref{fig:splus_smcclusters}  we show examples of some of the stellar clusters studied in this work, with combined images of the 12 bands using the \textit{S-PLUS cloud}  \footnote{\url{https://splus.cloud/}}. In this figure, it is clear that the resolved photometry of stars in the cluster cores is heavily limited by the seeing. In this paper, we chose to not compute the resolved photometry of stars not only due to confusion in the cores but also due to the depth of the S-PLUS survey (typically 20 AB mag in the narrow-band filters and 20.5 AB mag in the sloan-like filters). Besides that, the pixel scale of SPLUS is 0.55 arcsec pixel$^{-1}$ \citep{MendesdeOliveira2019}, therefore, even a better seeing would not allow a resolved photometric analysis of the crowded cluster cores as we can see in Figure \ref{fig:splus_smcclusters_zoom}, a zoom of the NGC 121 star cluster. S-PLUS images at the distance of the SMC are not suitable to reach the main sequence turnoff to fit an isochrone to the color-magnitude diagram of the stars.

\subsection{Stellar Clusters reference catalog}
\label{sec:bica}

Since the Magellanic Clouds are very close to our Galaxy, they have many stellar clusters with parameters well determined in the literature. A recent effort to put all the best determinations of these parameters for SMC clusters in a single catalog can be found in \citet{2020AJ....159...82B}. This list includes stellar clusters, associations, and extended objects related to stellar populations. We can, then, use this catalog to validate the method that uses integrated photometry of the clusters with S-PLUS and apply it to obtain new results. Only objects from the catalog classified as stellar clusters were used for this work. From this catalog, values of ages, metallicities (when available), spatial coordinates of the centers, and sizes were extracted. In cases where there was more than one reliable age determination, the catalog itself included the geometrical averages between the values, resulting in a single value for each object. As for metallicity, the catalog prioritized results following the order: calcium triplet and other high-resolution spectroscopy of red giants, isochrone fitting of CMDs, integrated spectroscopy, and integrated photometry taken from \cite{Bica1986A&A...156..261B}. This order was chosen to give priority to the most reliable methods of metallicity determination.

\subsection{Photometry using \texttt{photutils} package}
\label{subsec:photutils}

\begin{figure*}
    \includegraphics[width=0.8\textwidth]{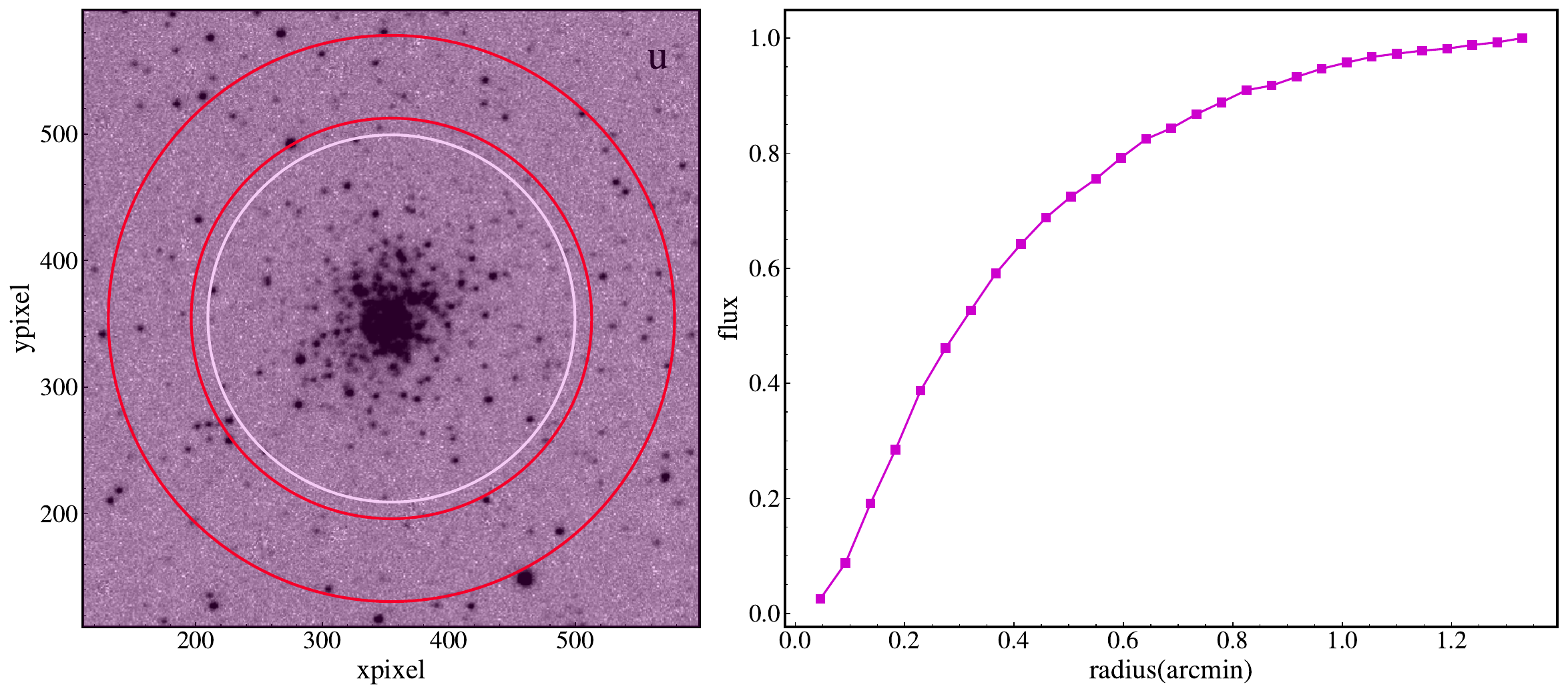}
    \includegraphics[width=0.8\textwidth]{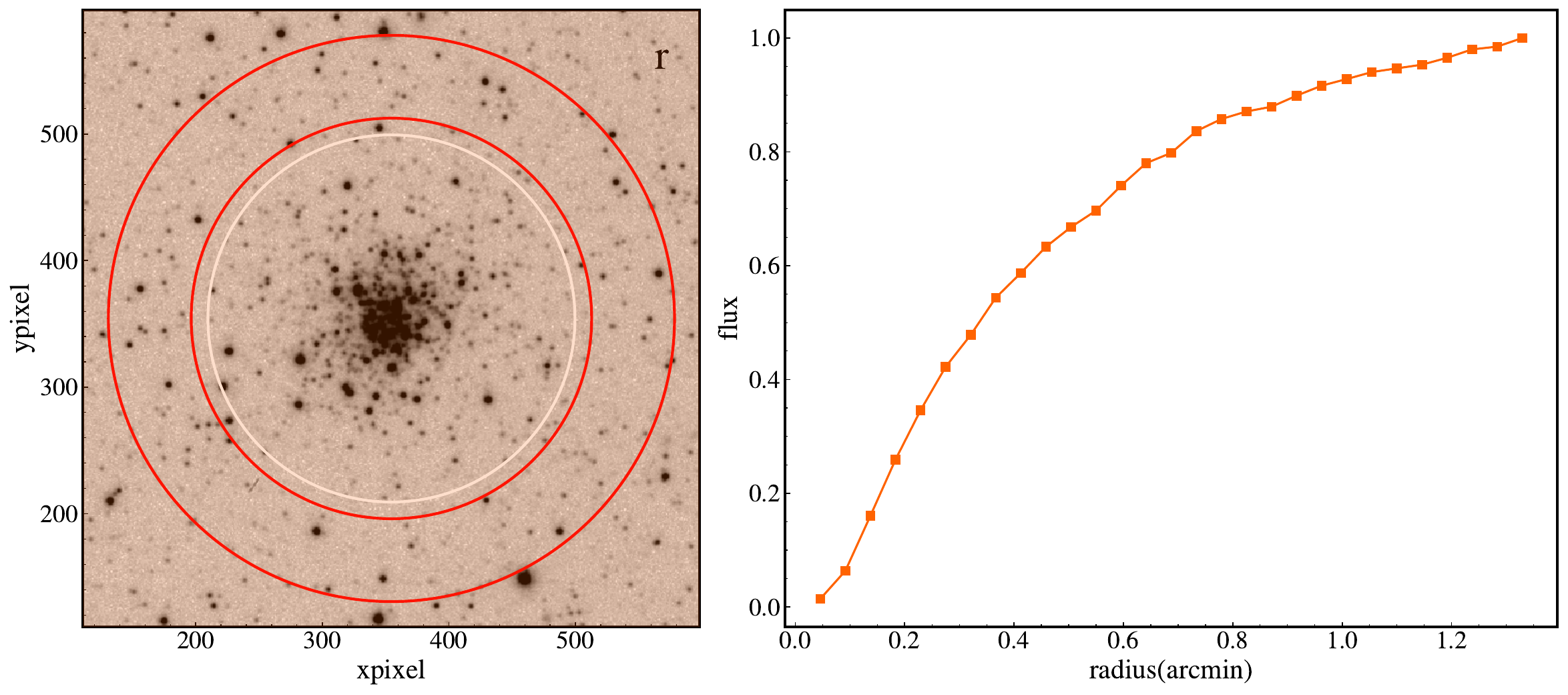}
    \caption{S-PLUS images of the \textit{u} (upper left panel) and \textit{r} (lower left) bands for the stellar cluster NGC 458. Red circles represent the ring used to obtain the sky flux level value, while white circles represent the aperture used to obtain the integrated flux of the cluster. The horizontal and vertical axes are the pixel positions of the image. In the right panels, we have the cumulative distribution of the flux as a function of the radius.}
    \label{fig:fluxcurve}
\end{figure*}

In order to obtain the integrated magnitudes of the clusters, we used the Python package \texttt{photutils} \citep{larry_bradley_2020_4044744}. This package allows obtaining the flux in circles centered on the stellar clusters. The radii were chosen from the maximum value between the two semi-axes, taken from \cite{2020AJ....159...82B}. An example of using the package is shown in Figure \ref{fig:fluxcurve}; white circles indicate the aperture used to obtain the cumulative flux of the object, while the outer ring of red circles represents the apertures chosen to obtain the background flux level, which must be subtracted from the measurement of the flux of the object. The inner radius of the background was chosen using the aperture of the cluster plus a value of $0.15$ arcminutes. For the width of the ring used to determine the background, we used an arbitrary fixed size of $0.6$ arcminutes for all clusters, which was considered a sufficient width to obtain a good measure of the background. To determine the background value, a median sigma clipping was used. Using this method to calculate the background value is important since it excludes outliers such as very bright stars in the field, meaning that only pixels with flux values within a range of 3$\sigma$ of the average value are kept. This is particularly important for images of the Magellanic Clouds since the fields in which the studied stellar clusters are located have a high concentration of objects. To obtain the final magnitude values, it was necessary to transform the fluxes into magnitudes. Zero points were obtained using the techniques in \citet{SPLUSDR2}, which describes the second data release of S-PLUS (SPLUS DR2). The extinction coefficients were calculated according to Section \ref{subsec:ext}. The reddening values used will be described in Subsection \ref{subsec:ebv}. The final equation used for the magnitude, corrected for zero point and extinction, is:

\begin{equation}
    mag_x = -2.5\times\log{F_x} + ZP_x - k_x \times E_{(B-V)},
    \label{eq:mag}
\end{equation}

\noindent where $mag_x$ is the final magnitude in each band $x$, $F_x$ is the total flux in $x$.
$ZP_x$ is the zero point for the given band in the S-PLUS field in which the cluster is located, obtained using the techniques described in \citet{SPLUSDR2}. The zero-point converts the measured instrumental magnitude to AB-calibrated magnitudes at the top of the atmosphere. This value accounts for the transformation to the AB-magnitude system, also taking into account the efficiency of the instruments, atmospheric conditions and the airmass at the time of the observation. The AB-calibrated magnitudes are still affected by the interstellar medium (ISM) extinction, which is corrected by subtracting $k_x \times E_(B-V)$ from the instrumental magnitudes. $k_x$ is the extinction coefficient, related to the S-PLUS filter system, calculated using the transmission curves of the survey's passbands. The color excess, $E_{(B-V)}$, is a measurement of the reddening in the direction of the cluster. In this work, we use the color excess maps provided by \citet{Gorski2020ApJ...889..179G}.

Throughout this work, the magnitude names $mag_x$ will be replaced with $x$, to indicate the magnitude in a given band. The uncertainty $\sigma_x$ of each magnitude was determined from 1\% of the magnitude value in the respective band. This choice was made considering all possible sources of errors: calibration and image. The faint end of our sample is $\approx17$ mag, so this turns out to be a pretty conservative error.

\subsection{ISM Extinction coefficients}
\label{subsec:ext}

\begin{table}
    \centering
    \caption{Extinction coefficients calculated for the 12-band filter system used by the S-PLUS, calculated using the extinction law from \citealt{Schlafly2016ApJ...821...78S}. }
    \label{table:kx}
    \begin{tabular}{c|c|c|c|}
        Filter name & $\lambda_c$ [$\angstrom$] & $\Delta\lambda$ [$\angstrom$] & $k_x$ \\
        $u$ & $3563$     & 352 & $4.90$ \\
        $J0378$ & $3770$ & 151 & $4.67$ \\
        $J0395$ & $3940$ & 103 & $4.48$ \\
        $J0410$ & $4094$ & 201 & $4.32$ \\
        $J0430$ & $4292$ & 201 & $4.11$ \\
        $g$ & $4751$     & 1545 & $3.67$ \\
        $J0515$ & $5133$ & 207 & $3.33$ \\
        $r$ & $6258$     & 1465 & $2.51$ \\
        $J0660$ & $6614$ & 147 & $2.30$ \\
        $i$ &  $7690$    & 1506 & $1.80$ \\
        $J0861$ & $8611$ & 408 & $1.46$ \\
        $z$ & $8831$     & 1182 & $1.39$\\ \hline

    \end{tabular}
\end{table}

As mentioned before, in order to correct the reddening of the clusters, the extinction coefficients ($k_x$) were calculated using the extinction law proposed by \cite{Schlafly2016ApJ...821...78S} with $R_v = 3.1 $. These coefficients were calculated using the central wavelength values of each of the S-PLUS bands, obtained from \cite{MendesdeOliveira2019}. The values used for the central wavelength ($\lambda_{c}$), FWHM, and $k_x$ are presented in Table \ref{table:kx}. Since the central wavelengths of the S-PLUS filter system are slightly different from those of the J-PLUS system, we also have a difference between the values of $k_x$ when comparing the values obtained in \cite{2019A&A...631A.119L}.

\subsection{Reddening maps}
\label{subsec:ebv}

The extinction coefficient of each band is associated with the reddening value $E(B-V)$ of the stellar cluster. Such an effect is caused by the interstellar dust of the Milky Way and the MCs themselves. For this work, the reddening maps of \cite{Gorski2020ApJ...889..179G} were used. The values were calculated from the difference between the intrinsic and the observed color in clumps of stars in the red giant branch. To assign reddening values to each of the studied clusters, we performed an interpolation of the maps as a function of R.A. and Decl., then calculated the reddening at the central coordinates.

\subsection{Inspection of the \textit{u} and \textit{r} bands}

Performing integrated photometry of stellar clusters in crowded fields can be a challenge. Problems, such as contamination by bright field stars that are not members of the cluster, led us to double-check the obtained photometry. For this purpose, we visually inspected the S-PLUS images in the \textit{u} and \textit{r} bands for all clusters. The flux growth curve of each cluster was analyzed in order to check if there was convergence to the total magnitude, or if there was contamination from neighboring sources. We excluded the cases where no convergence was reached. We present an example of this exercise's results in Figure \ref{fig:fluxcurve} for the cluster NGC 458.

\begin{figure}
    \includegraphics[width=1\columnwidth]{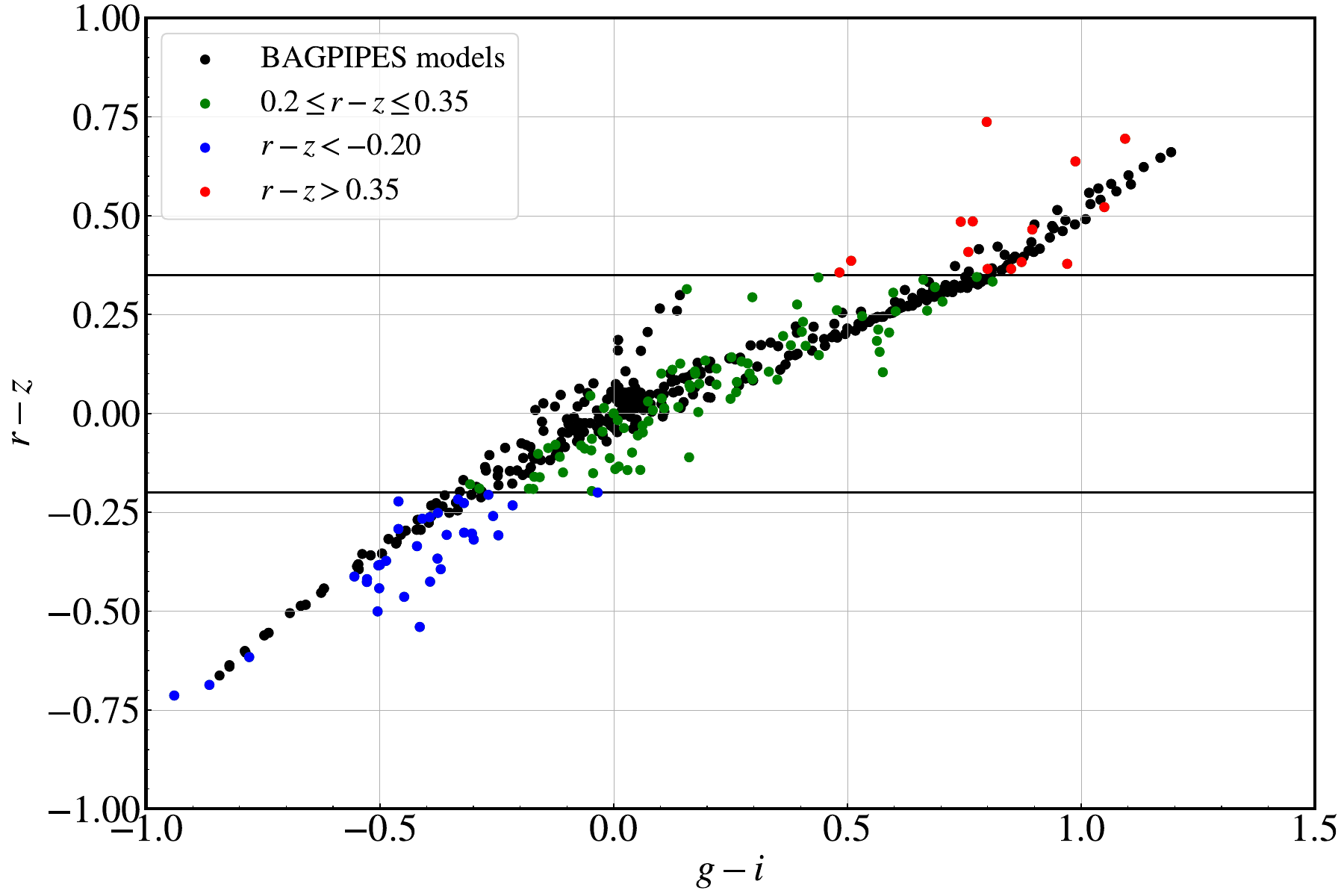}
    \caption{Simple stellar population models (SSPs) used in BAGPIPES (black dots). Red circles are all of our starting sample, while blue are the ones in the final sample analyzed in this work. In blue we have star clusters with $r-z < -0.20$, and, in red, $r-z > 0.35$. Green circles are stellar clusters with colors between these two values.}
    \label{fig:models}
\end{figure}

Figure \ref{fig:models} shows the sample after this visual inspection. The stellar population models used in this work were compared to our data. Black circles show the models used in \texttt{BAGPIPES} (see Section \ref{subsec:bagpipes} for more details), while the blue, green, and red circles represent the data. As can be seen in the figure, not all stellar clusters have a good agreement with the models. There are two main regions, one with small $r-z$ (${<}-0.20$, blue circles) and the other with higher values of $r-z$ (${>}+0.35$, red circles), with the largest number of clusters with photometry different from what was expected from the models.  There can be numerous reasons for the disagreement between the photometry and the models, such as the presence of Be stars or \hii\ emissions, which can cause the clusters to appear bluer, but we did not check any further. Therefore, we decided to exclude these two regions from our sample, being restricted to where the grid is densest (green clusters, where $-0.20 \leq r-z \leq 0.35$), and to avoid extrapolating beyond the model grid. After all the procedures, the sample of this work contains 88 stellar clusters, 11 with no age and 65 with no metallicity determinations from the \citet{2020AJ....159...82B} catalog.

\section{Methodology}
\label{sec:methods}

\subsection{\texttt{BAGPIPES}}
\label{subsec:bagpipes}

\begin{figure*}
    \centering
    \includegraphics[width=2\columnwidth]{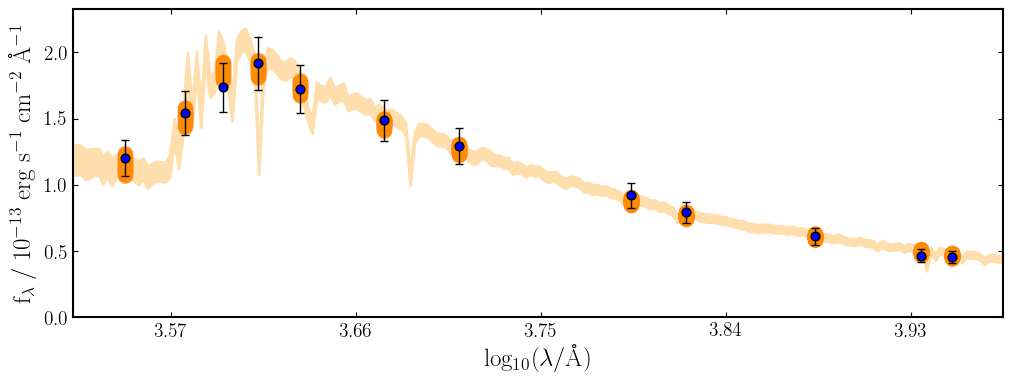}
    \caption{\texttt{BAGPIPES} fitting example for S-PLUS photometry for NGC 458 using a burst star formation history. Blue dots are the S-PLUS data with the respective errors. The shaded orange area represents the 16th and 84th percentile range for the posterior spectrum and photometry.}
    \label{fig:bagpipes_ngc458-fit}
\end{figure*}

\begin{figure}
    \centering
    \includegraphics[width=1\columnwidth]{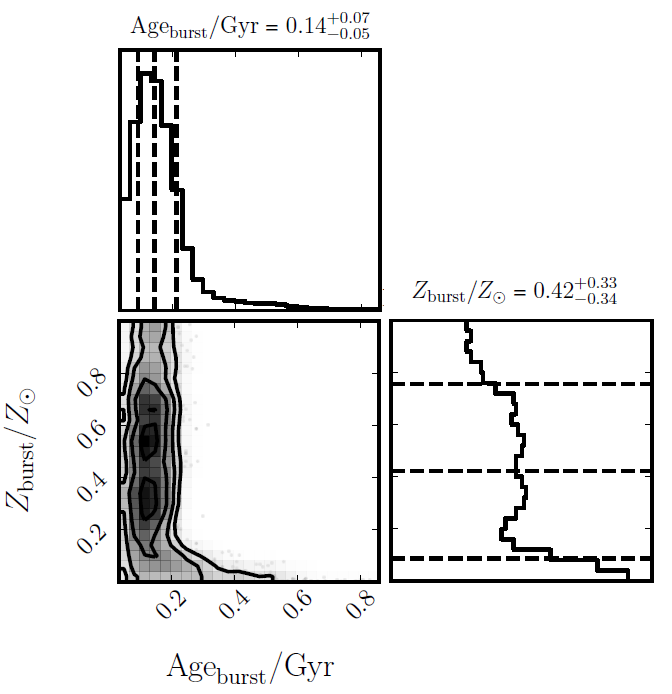}
    \caption{Corner plot for the \texttt{BAGPIPES} fitting example of the NGC 458 stellar cluster. The median, 16\% and 84\% values of \texttt{BAGPIPES} are also shown in the plot as dashed lines.}

    \label{fig:bagpipes_ngc458-corner}
\end{figure}

\texttt{BAGPIPES} is a code that allows one to create galaxy SEDs from parametric models of star formation histories, as well as adjust spectra and photometry for cases without sufficient resolution. The models include both the stellar continuum and the emission lines produced by the ionized gas from star-formation regions, the latter being estimated with the photo-ionization modeling software \texttt{Cloudy} \citep{Ferland2017}. Both the continuum and the emission lines are subject to attenuation by dust and gas present along the line of sight. Differential attenuation effect is also considered, in which young stars are immersed in their parent molecular cloud as well as behind a curtain of interstellar dust. Thus, young stellar populations are subject to more attenuation than old ones \citep{Charlot2000}. Being able to take all these parameters into account makes \texttt{BAGPIPES} a potentially great tool for analyzing unresolved stellar populations in S-PLUS as well as other similar surveys.

The code works under several assumptions that must be taken into consideration. To begin with, the spectrum of a galaxy can be modeled as a sum of simple stellar populations (SSPs), whose combination is directly dependent on the galaxy's star-formation history (SFH), in addition to the transmission functions of the ionized and neutral interstellar medium. From this, it is possible to find the galaxy luminosity, which, once calculated, is converted to flux density, using the distance and the transmission function of the intergalactic medium (see \citealt{Conroy2013ARA&A..51..393C} for a review). The SSPs used in this code are from the 2016 version of \citet{Bruzual2003} constructed using an initial mass function from \citet{KB2002.2002MNRAS.336.1188K}. For the SFH, we account for the stellar mass formed for each period during the SFH. For this task, we considered the full range of possible choices provided by \texttt{BAGPIPES}, but always with one parametric form of SFH behavior. Since we are using a galaxy-fitting code for stellar clusters, we adopted a single burst SFH considering that all stars of the cluster were formed at the same time with the same metallicity.

\begin{table}
    \centering
    \caption{List of free and fixed parameters. To the left, we have the free parameters (uniform priors), where $M_{\rm formed}$ is the stellar mass formed, $\rm age$ is the age of the stellar population, $Z$ is the metallicity, and $z$ is the redshift. The fixed parameters are the ionization parameter, $\log_{10}(U)$, and the ones related to dust attenuation modeling, $A_V$ and $\eta$ (both are zero because we are already derredening the photometry).}
    \begin{tabular}{|ll|ll|}
        \hline
        Free Parameter & Limits & Fixed Parameter & Value  \\ \hline
         $\log_{10}$($M_{\mathrm{formed}}$ / $M_{\odot}$) & $(0,13)$ & $\eta$ & $0$\\ 
         age / Gyr & $(0.01,15)$  & $A_V$ & $0$ \\
         $Z$ / $Z_{\odot}$ & $(0,1)$ & \\
         $z$ & $(0,0.001)$ & \\ \hline
    \end{tabular}
    
    \label{tab:bp_param}
\end{table}

All of these parameters are taken into account to perform the fitting: The code uses Bayesian inference, using as priors the stellar population parameters and other free parameters, checking and updating their probabilities at each realization. In Table \ref{tab:bp_param}, we have the free and fixed parameters used in this work with the upper limit for age being 15 Gyr. The redshift was set in the range of the known SMC redshift \citep{SMCz.2012AJ....144....4M}, from 0 to 0.001, and the mass range from 0 to 13 ($\log(M/M_{\odot})$).  We tested the mass range and found that it has no influence on the determination of the metallicities and ages. In fact, mass and distance will not change the shape of the spectrum and therefore will not affect the derived parameters. Lastly, all the dust-related parameters were fixed as zero since the clusters were already dereddened. Figures \ref{fig:bagpipes_ngc458-fit} and \ref{fig:bagpipes_ngc458-corner} represent the \texttt{BAGPIPES} results for NGC 458,  confirming the correlation between age and metallicity seen in the literature (e.g. \citep{Bica1986A&A...156..261B}).

\section{Results}
\label{sec:results}

\subsection{Comparison with the literature}

To apply the method to new stellar clusters, it was necessary to validate the method for clusters with previously known parameters. For this, we compared the results of this work with results from \cite{2020AJ....159...82B} for clusters with known age and metallicity. The metallicity values from \texttt{BAGPIPES} have the format $\frac{Z}{Z_{\odot}}$ with $Z_{\odot}=0.02$, where $Z_\odot$ is the mass-fraction solar metallicity. To compare to most of our literature, we are multiplying the output value of the code by $\frac{0.020}{0.0152} \approx 1.316$, since we are adopting $Z_{\odot}=0.0152$. We approximate the [Fe/H] values using [Fe/H]$\equiv\log(\frac{Z}{Z_{\odot}})$. 

\begin{figure*}
\begin{center}
\includegraphics[width=\textwidth]{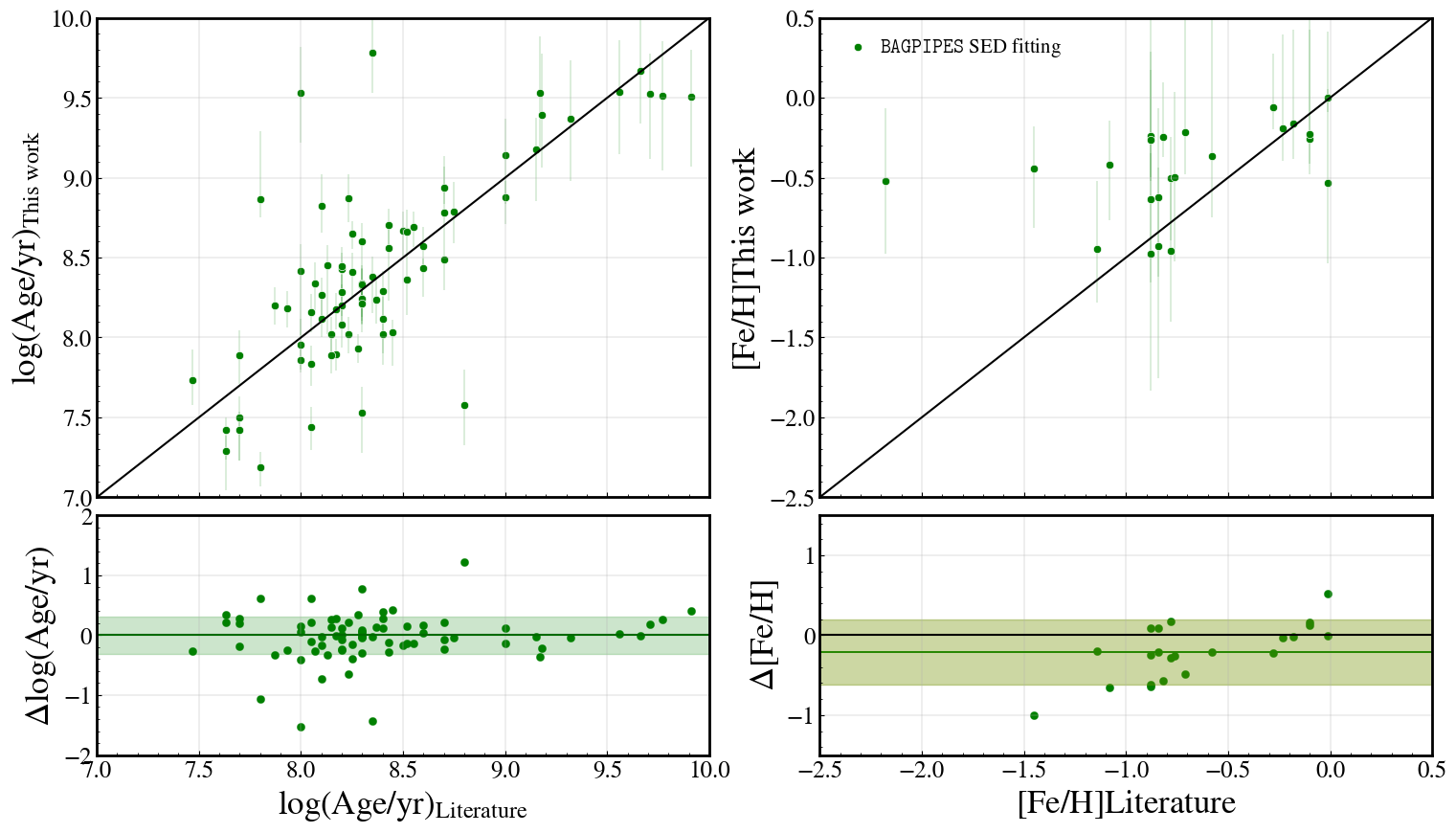}

\caption{The top panels show the comparisons between the values derived in this work and the literature values for age (left) and metallicity (right). The bottom panels are the residuals in age and metallicity between the literature and derived values, with the shaded area being the dispersion of the values. The dispersion values were 0.31 for $\log(\rm age)$ and 0.41 dex for [Fe/H].}
 \label{fig:litcomparison_bestsample}
\end{center}
\end{figure*}

In Figure \ref{fig:litcomparison_bestsample}, we have our clusters with ages and metallicities already known and we compare the literature values with the ones obtained using \texttt{BAGPIPES}. For age, there is a good agreement with the reference, with a larger dispersion for clusters with age $< 10^{8.6}$ yr when compared to the rest of the test sample. We encountered a residual when compared to the literature values of around 0.31 for $\log(age)$ and 0.41 dex for [Fe/H]. These values were calculated using a bi-weight scale \citep{Beers1990AJ....100...32B}, which diminishes the effects of outliers. From the results obtained with \texttt{BAGPIPES}, metallicities comparable to those obtained through low-resolution spectroscopy of individual stars were determined ($\gtrsim0.2$ dex, \citealt{Yanny2009AJ....137.4377Y}). In the case of metallicity, it is also clear that there is an offset of $0.21$ dex, which means that \texttt{BAGPIPES} is overestimating the metallicities.  Even though we are not able to derive accurate metallicities for individual clusters,  patterns regarding the metallicity distribution of the SMC  can still be detected.
\begin{figure*}
\begin{center}
\includegraphics[width=\textwidth]{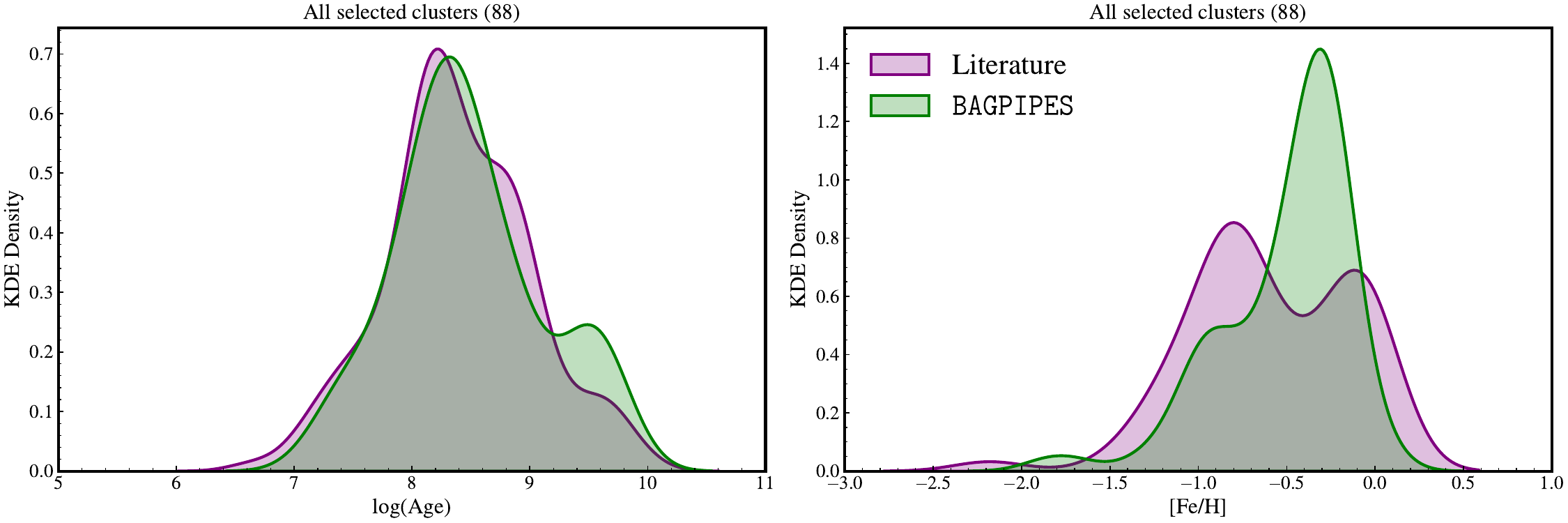}

\caption{KDE age and metallicity distributions. \texttt{BAGPIPES} distributions are represented in green, and literature values are in purple. On the left, we have the age distributions, while on the right we have the metallicity distributions.}
 \label{fig:kde_final_best}
\end{center}
\end{figure*}

We can also assess the distributions of ages and metallicities to verify if our obtained parameters are in agreement with the literature. The two panels of Figure \ref{fig:kde_final_best} show the probability density distributions using kernel density estimators (KDEs). The age distributions of our work have a first peak near $\log(age)=8.5$, which is close to the literature, and a second peak $\log(age)\sim 9.5$. As for metallicity in \texttt{BAGPIPES}, we have a much higher concentration of clusters with [Fe/H] around $-0.4$ and a second less pronounced peak at [Fe/H]$=-1.0$. The peak at [Fe/H]$=-1.0$ is similar to the peak found in the literature.

To account for the number of peaks the age and metallicity distributions have and their positions, we can fit Gaussian Mixture Models \citep[GMM,][]{Muratov2010ApJ...718.1266M} to the distributions of ages and metallicities obtained. After testing different numbers of components for each parameter, we selected the model with the lowest Bayesian Information Criterion  \citep{Schwarz1978.10.1214/aos/1176344136},  value as the best GMM fitted. In the case of age, the bimodal distribution with peaks at $\log(\rm age) = 8.3\pm0.5$ and $\log(\rm age) = 9.6\pm0.1$ was better fitted, which is linked to the large number of young clusters found in the center of the Small Magellanic Cloud. For metallicity, the best fit was found using the bimodal distribution, with peaks at
$-0.3\pm0.1$ dex and $-0.8\pm0.4$ dex. These fits were also done for the literature sample with 314 ages and 62 metallicities, resulting in an unimodal distribution for the age ($8.4\pm0.6$) and a bimodal for the metallicity ($-0.1\pm0.1$ dex and $-0.8\pm0.4$ dex). 

Recent literature on stellar populations in the SMC indicates that there is a metallicity peak at $-0.7$ dex for young populations \citep{Karakas2018MNRAS.477..421K} and between $-0.8$ dex and $-1.0$ dex a peak for red giants \citep{D'Onghia2016ARA&A..54..363D}. We can also compare the unimodal and bimodal metallicity peaks found with those in \cite{Parisi2022A&A...662A..75P}, obtained through the spectroscopy of red giant stars using the Calcium Triplet. This work states that the greater the number of stellar clusters, the less significant the bimodality is, compared to \citet{PGC+15.2015AJ....149..154P}, where there were fewer stellar clusters in the sample. Later, \citet{DeBortoli2022A&A...664A.168D} showed that this possible bimodality happens when we consider two different regions of the SMC (inner and outer region). Although \texttt{BAGPIPES} results provided a bimodal shape to the metallicity distribution, one of the peaks is much higher
than expected.

\subsection{Spatial distributions}

\begin{figure}
\begin{center}
\includegraphics[width=1\columnwidth]{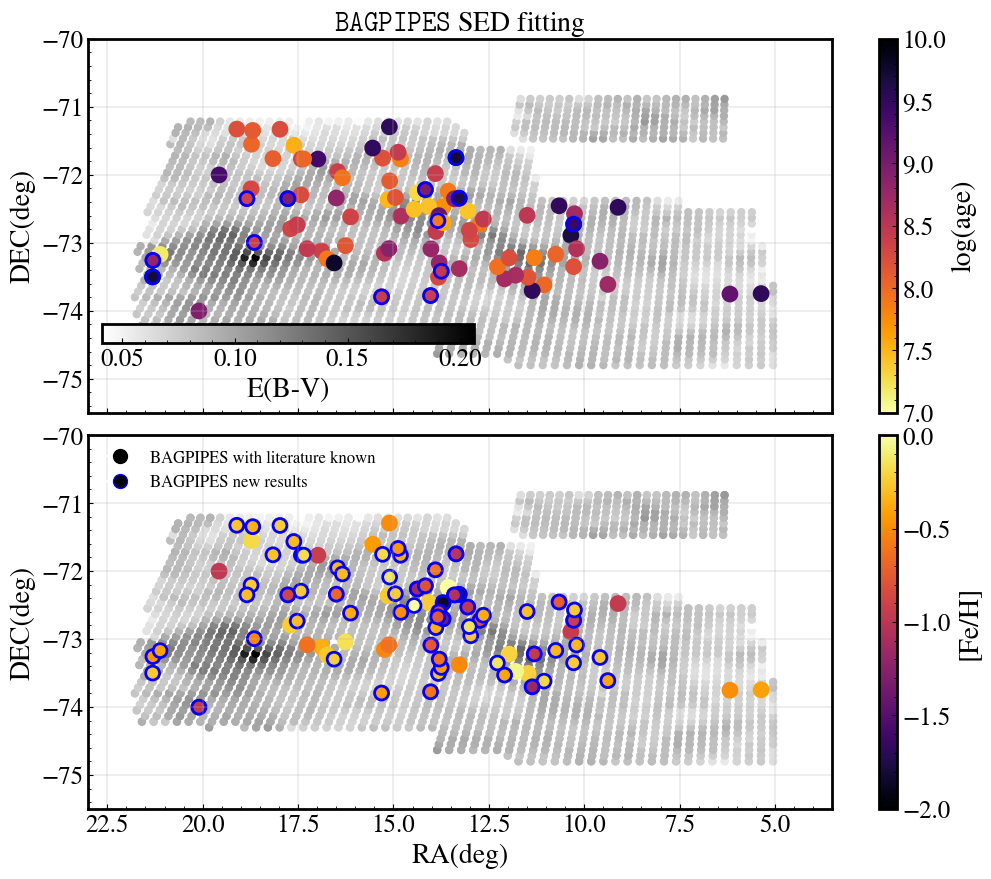}

\caption{Spatial distribution of ages and metallicities (colorbars in the upper and lower panels, respectively) using \texttt{BAGPIPES}. Circles with no edges are stellar clusters that have values in the literature, while those with blue edges are new values.}
\label{fig:map_final_best}
\end{center}
\end{figure}
We, now, look at the derived ages and metallicities as a function of the spatial coordinates (Figure \ref{fig:map_final_best}). We found that the core regions of this galaxy are predominantly occupied by young clusters, while its outskirts contain mostly older ones. This result is, perhaps, not unexpected given that the young clusters are associated with star-forming regions found in the center of SMC. However, we now confirm this expectation with a larger sample of clusters with homogeneously derived parameters than ever before. We also note the existence of some young clusters at the edges of the SMC associated with the Magellanic Bridge \citep{Piatti2015MNRAS.450..552P}.  In order to quantify this observation, we analyzed the evolution of the R.A. and Decl. dispersions in bins of $\log{(\rm age)}$ (Table \ref{tab:grad_age_best}). We found 1$\sigma$ evidence for older clusters ($\log{(age)} \geq 9$) being more spatially diffuse than young ones ($\log{(\rm age)} < 9$). Interestingly, despite this clear existence of an age gradient, we found no evidence for a metallicity gradient from our cluster data in the SMC (Table \ref{tab:grad_feh_best}). Indeed, the metallicity gradient in the SMC is estimated to be quite small,  at the level of $-0.031\pm0.005$ dex/kpc \citep{Choudhury2018MNRAS.475.4279C} and our BAGPIPES values might not be precise enough to probe it.

\begin{table}
\centering
\caption{Standard deviations of RA and DEC for stellar clusters divided into 5 age groups in order to verify if there is a gradient towards the center of the galaxy.}
\label{tab:grad_age_best}
\begin{tabular}{|l|l|l|l|}
\hline
\texttt{BAGPIPES} & $\sigma_{RA}$ (deg)    & $\sigma_{DEC}$ (deg)   & N \\ \hline
$\log(age)\geq$ 9.0          & 4.30 & 0.78 & 16  \\
8.5$\leq\log(age)<$ 9.0      & 3.28 & 0.50 & 18 \\
8.0$\leq\log(age)<$ 8.5      & 2.62 & 0.73 & 37 \\
7.2$\leq\log(age)<$ 8.0      & 1.59 & 0.48 & 16 \\ \hline

\end{tabular}
\end{table}

\begin{table}
\centering
\caption{Standard deviations of the RA and DEC for 3 different groups of metallicity.}

\label{tab:grad_feh_best}
\begin{tabular}{|l|l|l|l|}
\hline
\texttt{BAGPIPES} & $\sigma_{RA}$ (deg)    & $\sigma_{DEC}$ (deg)   & N \\ \hline
[Fe/H] $> -0.4$           & 2.93 & 0.71 & 45 \\
$-0.8 < $ [Fe/H] $ \leq -0.4$ & 3.54 & 0.66 & 26 \\ 
$-1.2<$[Fe/H] $ \leq -0.8$ & 3.37 & 0.65 & 14 \\ \hline
\end{tabular}
\end{table}

Another way we can investigate these gradients is by relating the stellar parameters to the projected distance of a given stellar cluster. This projected distance is the semi-major axis of the ellipse around the center of the SMC, concentric to the LMC main body outline, which passes through the cluster
\citep{Piatti2007MNRAS.377..300P, Dias2014A&A...561A.106D, DKB+16.2016A&A...591A..11D, PGC+16.2016AJ....152...58P}. Then, we can compare directly with Figure 8 from \cite{2020AJ....159...82B}. As we have stellar clusters only at $\lesssim3$ degrees from the SMC's cen, they are mostly contained within the main body of the SMC, thus we do not see any inversion in the correlation between the parameters and the projected distance to the center, as it only happens at a$~4$-$5$ degrees. However, we were still able to confirm that the age/metallicity variations prior to $~3$ degrees are not large both in age and metallicity.

\subsection{Age-metallicity relation}

\begin{figure*}
\begin{center}
\includegraphics[width=1\textwidth]{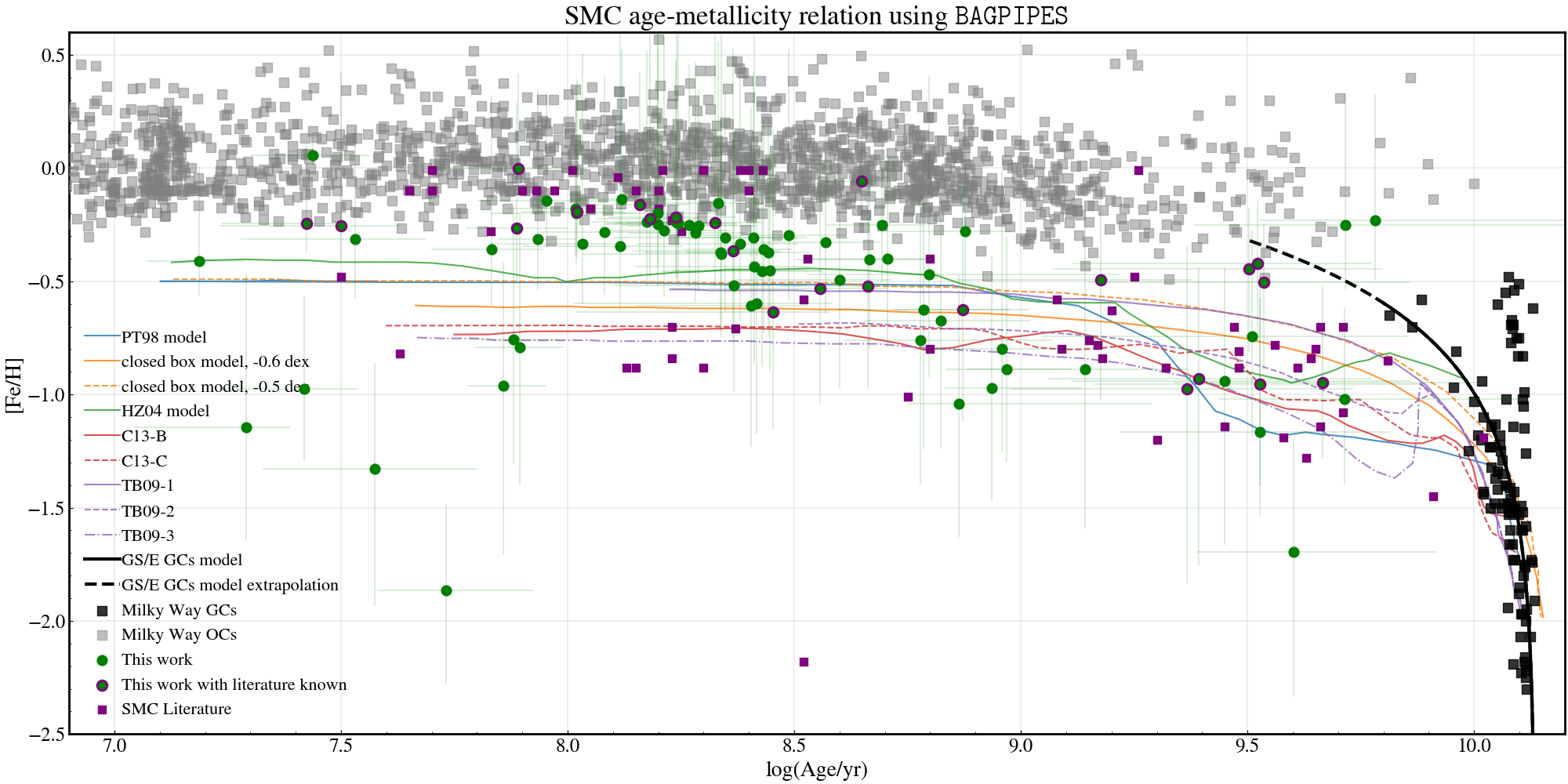}
\caption{AMR using \texttt{BAGPIPES}. In green, we have the results for all of our sample. Those with a purple outline, are clusters with known parameters in the literature. Purple squares are the literature values for the SMC clusters found in \citealt{2020AJ....159...82B} that are within the S-PLUS fields. Theoretical models are also present in the plot: PT98 "burst" model (blue line); closed-box models for [Fe/H]$=-0.6$ dex and [Fe/H]$=-0.5$ dex (full and dashed orange lines, respectively); the HZ04 model (green line); models C13-B and C13-C (full and dashed red lines, respectively); models TB09-1, TB09-2, and TB09-3 (full, dashed and dot-dashed purple lines, respectively). To compare with our Galaxy, we also have a sample of globular clusters  from \citealt{Kruijssen2019MNRAS.486.3180K}  (black squares), a sample of open clusters from \citealt{Dias2021MNRAS.504..356D}  (gray squares), and the model for the Gaia-Sausage/Enceladus accreted by the MW from \citealt{Limberg2022ApJ...935..109L} (full black thick line) along with an extrapolation (dashed black thick line)}
 \label{fig:amr_bagpipes}
\end{center}
\end{figure*}

The age-metallicity relation (AMR) of our sample is shown in Figure \ref{fig:amr_bagpipes}. The different lines show different chemical-evolution models known from the literature: PT98 \citep[burst model, blue line]{PT1998MNRAS.299..535P}; closed-box models \citep[orange lines]{DH98.1998AJ....115.1934D}; model HZ04 \citep[green line]{HZ04.2004AJ....127.1531H}, models C13-B and C13-C \citep[red lines,][]{C2013ApJ...775...83C} are the two Bologna models; TB09-1, TB09-2 and TB09-3 models \citep[purple lines][]{TB09.2009ApJ...700L..69T}. Comparing the theoretical models, the literature values for the clusters, and our results, we can also see that the ones from our sample closely follow both the literature age-metallicity relationship and the theoretical models, but the results are better for stellar clusters with $\log(age)>8.0$. 

With this new set of parameters for SMC stellar clusters at hand, we are also able to compare it with the AMR of our Galaxy. For this, we utilize the sample of globular clusters (GCs) from \citet{Kruijssen2019MNRAS.486.3180K}. This compilation includes both in-situ GCs, born in the Milky Way itself, as well as ex-situ ones, which were accreted over the course of the Galaxy's hierarchical assembly history (e.g., \citealt{Massari2019}). The most prominent ex-situ population consists of GCs brought into the Milky Way's halo alongside the infall of the Gaia-Sausage/Enceladus (GSE; \citealt{Belokurov2018}, \citealt{Haywood2018}, and \citealt{Helmi2018}), the dwarf galaxy associated with the last significant merger experienced by the Galaxy. Additionally, we compare our results to ages and metallicities of Galactic open clusters (gray squares) from \citet{Dias2021MNRAS.504..356D}.

The most note-worthy feature of Figure \ref{fig:amr_bagpipes} is that, at a given age, our results predict lower metallicities in comparison to the GSE model (black line) of \citet[][see also \citealt{Forbes2020} and \citealt{Callingham2022}]{Limberg2022ApJ...935..109L}. This behavior is naturally explained by means of the mass-metallicity relation (e.g., \citealt{Kirby2013ApJ...779..102K}), given that the GSE progenitor is expected to be more massive than the SMC \citep{Feuillet2020, Naidu2020, Naidu2021} and that the SMC has been much less efficient in star formation as expected for a galaxy with this mass \citep{Nidever2020ApJ...895...88N}. Moreover, this result is also consistent with chemical-evolution studies based on stellar abundances for both the GSE and the
SMC \citep{Hasselquist2021}. The obvious limitation regarding this conclusion is that our results account for all star clusters in our S-PLUS SMC sample, while the available GSE models available can only account for old GCs. However, in comparison with Galactic open clusters, SMC ones are also more metal-poor at any given age. Therefore, it is reassuring that our result is consistent with expectations from standard galaxy-evolution assumptions and well-known previous literature models of the SMC.

The age-metallicity relation of this work also corroborates with the results found in \citet{Oliveira2023MNRAS.524.2244O}  from clusters towards the SMC Wing/Bridge. Both studies observed a slight metallicity dip near a $\log({\rm age})=9$, which they related to a consequence of the same galaxy-galaxy interaction that generated the Magellanic Stream around that time. Furthermore, the Wing/Bridge was formed of material pulled from the SMC main body around $\sim150$~Myr ago (e.g. \citet{Zivick2018ApJ...864...55Z}). Most of the clusters analyzed in this work are located in the central region of the SMC. Therefore, the fact that the chemical evolution of the Main Body from the present work and that of the Wing/Bridge from \citet{Oliveira2023MNRAS.524.2244O} are similar, is consistent with the formation scenario of the Magellanic Bridge. Recently,  \citet{Almeida2023arXiv230813631A} analyzed the chemical abundances of field stars and found a similar conclusion in the sense that the chemical evolution and metallicity distribution of the main body and bridge are similar. In summary, the results from this work are revealing the chemical evolution of the host galaxy when we analyze the star clusters as a population, even if individual cluster ages and metallicities have larger uncertainties with respect to CMD results for example, and
prove the power of this technique and of the S-PLUS multi-wavelength photometry.

\section{Summary and closing comments}
\label{conclusao}

In this work, we present a method to estimate the ages and metallicities of SMC stellar clusters from integrated photometry of S-PLUS. We apply the new methodology to the SMC as a test case. In the future, the technique developed here might be applied to other external galaxies even outside the local group. The color interval $-0.20 \leq r - z \leq 0.35$ was defined as the one with the most reliable density of models to cover all of our data. This interval contains 88 SMC clusters, 11 with ages and 65 with metallicities not previously determined in the literature.

\begin{itemize}

\item We obtained residuals compared to literature values with \texttt{BAGPIPES} at the level of $\Delta \log(age)=0.31$ for age and $\Delta$[Fe/H]$ = 0.41$ dex for metallicity.

\item The distribution of ages obtained was compatible with the literature distribution, with a peak for ages of $\sim$100 Myr. However, our \texttt{BAGPIPES} metallicities had a higher concentration of clusters at [Fe/H] $\approx -0.5$ and a smaller peak near [Fe/H]$\sim-0.9$. We utilized a GMM method to identify one single peak for ages, at $\log(age)=8.6\pm0.6$, related to a large number of young clusters in the sample. For metallicities, we found a more metal-rich peak at [Fe/H]$=-0.4\pm0.2$ and a more metal-poor one at [Fe/H]$=-0.9\pm0.2$.

%age gradients and metallicities
\item  We managed to reproduce the previously known projected age gradient at the 1$\sigma$ level, where older clusters are in the outermost regions of the galaxy and younger clusters are closer to the center. On the other hand, we were not able to find any significant projected metallicity gradient. 

\item We were able to compare our stellar clusters with the ex-situ Milky Way GCs that stem from the GSE dwarf galaxy. When compared to the AMR from the GSE, we confirmed that the SMC contains stellar clusters with lower metallicity values, as expected given the difference between the masses of each galaxy and the known lower star formation efficiency of the SMC.

\item  Our research findings are in line with the discoveries made by other studies regarding the correlation between age and metallicity in SMC Wing/Bridge. It indicates a small dip in metallicity at a log(age) of 9, which confirms consistency between the Wing/Bridge and the central regions of SMC, which is consistent with the formation scenario of the Magellanic Bridge.
\end{itemize}

The strength of our method is not to derive accurate parameters for individual clusters, our main goal is, instead, to detect patterns in the stellar age and metallicity distribution throughout the SMC by analyzing the star clusters as a population. Although parameters estimated from the tools presented here still carry significant uncertainties, we were able to provide, for a large sample and in a homogeneous way, ages and metallicities for SMC clusters and provide constraints to the SMC chemical and dynamical evolution history. As they are important to study the formation history of galaxies through stellar clusters, the methods presented in this paper can be applied to stellar clusters in other galaxies both within and outside the local group in S-PLUS as well as other similar surveys.
\section*{Acknowledgements}

%The Acknowledgements section is not numbered. Here you can thank helpful
%colleagues, acknowledge funding agencies, telescopes and facilities used etc.
%Try to keep it short.

G.F.S. acknowledges CAPES (procs. 88887.481161/2020-00 and 88887.707003/2022-00).
F.A.-F. acknowledges funding for this work from FAPESP grants 2018/20977-2 and 2021/09468-1. G.L. acknowledges FAPESP (procs. 2021/10429-0 and 2022/07301-5). B.D. acknowledges support by ANID-FONDECYT iniciación grant No. 11221366. E. M.-P., acknowledges CAPES (proc. 88887.605761/2021-00). R.G. acknowledges FAPERJ (procs. E-26/205.964/2022 and 205.965/2022). 

The S-PLUS project, including the T80-South robotic telescope and the S-PLUS scientific survey, was founded as a partnership between the Funda\c{c}\~{a}o de Amparo \`{a} Pesquisa do Estado de S\~{a}o Paulo (FAPESP), the Observat\'{o}rio Nacional (ON), the Federal University of Sergipe (UFS), and the Federal University of Santa Catarina (UFSC), with important financial and practical contributions from other collaborating institutes in Brazil, Chile (Universidad de La Serena), and Spain (Centro de Estudios de F\'{\i}sica del Cosmos de Arag\'{o}n, CEFCA). We further acknowledge financial support from the S\~ao Paulo Research Foundation (FAPESP), Funda\c{c}\~ao de Amparo \`a Pesquisa do Estado do RS (FAPERGS), the Brazilian National Research Council (CNPq), the Coordination for the Improvement of Higher Education Personnel (CAPES), the Carlos Chagas Filho Rio de Janeiro State Research Foundation (FAPERJ), and the Brazilian Innovation Agency (FINEP). The authors who are members of the S-PLUS collaboration are grateful for the contributions from CTIO staff in helping in the construction, commissioning and maintenance of the T80-South telescope and camera. We are also indebted to Rene Laporte and INPE, as well as Keith Taylor, for their important contributions to the project. From CEFCA, we particularly would like to thank Antonio Mar\'{i}n-Franch for his invaluable contributions in the early phases of the project, David Crist{\'o}bal-Hornillos and his team for their help with the installation of the data reduction package \textsc{jype} version 0.9.9, C\'{e}sar \'{I}\~{n}iguez for providing 2D measurements of the filter transmissions, and all other staff members for their support with various aspects of the project.

\section*{Data Availability}

The final photometric quantities measured are provided in a Zenodo repository: \url{https://doi.org/10.5281/zenodo.8384511}. The raw S-PLUS images are available upon request to the corresponding author. Nevertheless, the complete S-PLUS data will become publicly available alongside the next data release, scheduled for 2023.

%%%%%%%%%%%%%%%%%%%%%%%%%%%%%%%%%%%%%%%%%%%%%%%%%%

%%%%%%%%%%%%%%%%%%%% REFERENCES %%%%%%%%%%%%%%%%%%

% The best way to enter references is to use BibTeX:

\bibliographystyle{mnras}
\bibliography{core} % if your bibtex file is called example.bib

%%%%%%%%%%%%%%%%%%%%%%%%%%%%%%%%%%%%%%%%%%%%%%%%%%

%%%%%%%%%%%%%%%%% APPENDICES %%%%%%%%%%%%%%%%%%%%%

%\input{sections/8-appendix}

%%%%%%%%%%%%%%%%%%%%%%%%%%%%%%%%%%%%%%%%%%%%%%%%%%

% Don't change these lines
\bsp	% typesetting comment
\label{lastpage}
\end{document}